\documentclass{sig-alt-release2}
\usepackage{graphicx}

\usepackage[]{hyperref}
\hypersetup{
	bookmarks=true,         
	unicode=false,          
	pdftoolbar=true,        
	pdfmenubar=true,        
	pdffitwindow=false,     
	pdftitle={Model-Driven Architectural Monitoring and Adaptation for Autonomic Systems},
	pdfauthor={Thomas Vogel and Stefan Neumann and Stephan Hildebrandt and Holger Giese and Basil Becker},
	pdfsubject={},
	pdfcreator={},
	pdfproducer={},
	pdfkeywords={autonomic computing, model-driven engineering, model synchronization, model transformation, software architecture},
	pdfnewwindow=true,
}

\global\boilerplate={
	\copyright\,Authors/ACM,\,2010. 
	This\,is\,the\,author's\,version of the work. It is posted here for your personal use. Not for redistribution. The definitive Version of Record was published in 
	{\scriptsize \url{https://doi.org/10.1145/1555228.1555249}}.
}

\begin{document}

\conferenceinfo{ICAC'09,} {June 15--19, 2009, Barcelona, Spain.} 
\CopyrightYear{2009}
\crdata{978-1-60558-564-2/09/06} 

\title{Model-Driven Architectural Monitoring and Adaptation for Autonomic Systems}

\numberofauthors{1} 

\author{
\alignauthor
Thomas Vogel, Stefan Neumann, Stephan Hildebrandt, Holger Giese and Basil Becker\\
       \affaddr{Hasso Plattner Institute at the University of Potsdam}\\
       \affaddr{Professor-Doktor-Helmert-Str. 2-3}\\
       \affaddr{14482 Potsdam, Germany}\\
       \email{$\{$prename$\}$.$\{$surname$\}$@hpi.uni-potsdam.de}
}

\maketitle

\begin{abstract}
Architectural monitoring and adaptation allows self-ma\-nage\-ment capabilities of autonomic systems to realize more powerful adaptation steps, which observe and adjust not only parameters but also the software architecture. However, monitoring as well as adaptation of the architecture of a running system in addition to the parameters are considerably more complex and only rather limited and costly solutions are available today.
In this paper we propose a model-driven approach to ease the development of architectural monitoring and adaptation for autonomic systems. Using meta models and model transformation techniques, we were able to realize an incremental synchronization between the run-time system and models for different self-management activities. The synchronization might be triggered when needed and therefore the activities can operate concurrently.
\end{abstract}
\vspace{-0.15cm}
\vspace{-0.15cm}
\vspace{-0.14cm}
%
\section{Introduction}\label{sec:intro}\noindent
Software has to be frequently adapted to changes in the environment to keep its value for the user \cite{Lehman98}. Software can be adapted by modifying program variables (\emph{parameter adaptation}) or by exchanging algorithmic or structural system components (\emph{compositional adaptation}) possibly changing the software architecture \cite{McKinley+2004}.  
Continuous adaptations are impeded by the complexity of today's software systems.
The vision of \emph{Autonomic Computing} (AC) approaches this problem by borrowing the concept of control loops, which originates from the domain of control engineering, and adapting it to suit business computing by self-management capabilities such as \emph{self-configuration}, \emph{self-healing}, \emph{self-optimization} and \emph{self-protection} \cite{Kephart&Chess2003}.
Most of the work on AC employs parameter adaptation and therefore can employ well understood control engineering techniques \cite{hellerstein:control}. For some systems this is sufficient, but sometimes compositional adaptations have to be employed to achieve the needed self-management goals \cite{KramerM2007}.
Therefore, each capability requires its own corresponding abstract view on a managed software system that reflects the run-time state of a system regarding architectural components, links and parameters as far as they are relevant for the capability. These views should be provided by models.
While there are many examples for the benefits of adapting architectures at run-time
(cf. \cite{GarCHSS04,OreizyTMW99,Yang2009319}), the model-driven development using meta-models of the architecture or other elements of autonomic systems such as policies has only recently found more attention \cite{10.1109/ITNG.2008.247,Solomon+2008,ChengZ06}.
The focus of these model-driven approaches is to exploit models for facilitating control loop activities. In this context, only first ideas to use model transformation techniques exist \cite{BeckerGiese08c,Song+2008}, but no efficient solution working online has been presented so far.

In this paper we propose a model-driven approach to ease the development of architectural monitoring and adaptation for autonomic systems. By employing model-driven engineering techniques and our optimized model transformation technique \cite{GieseHildebrandt08, GW2008} the run-time system and several models aiming at different self-management capabilities are synchronized online and incrementally. The approach has been evaluated by means of an example implementation for \emph{Enterprise Java Beans}~\cite{jsr220}, which supports only the monitoring part yet. It considers performance monitoring, architectural constraints checking and failure monitoring.
The next section outlines our model-driven approach and the implementation, which is followed by a conclusion and an outlook on future work.
\section{Approach}\label{sec:approach}
\begin{figure}[htb]
        \centering
        \scalebox{0.7}{
        \includegraphics[keepaspectratio, width=1\columnwidth]{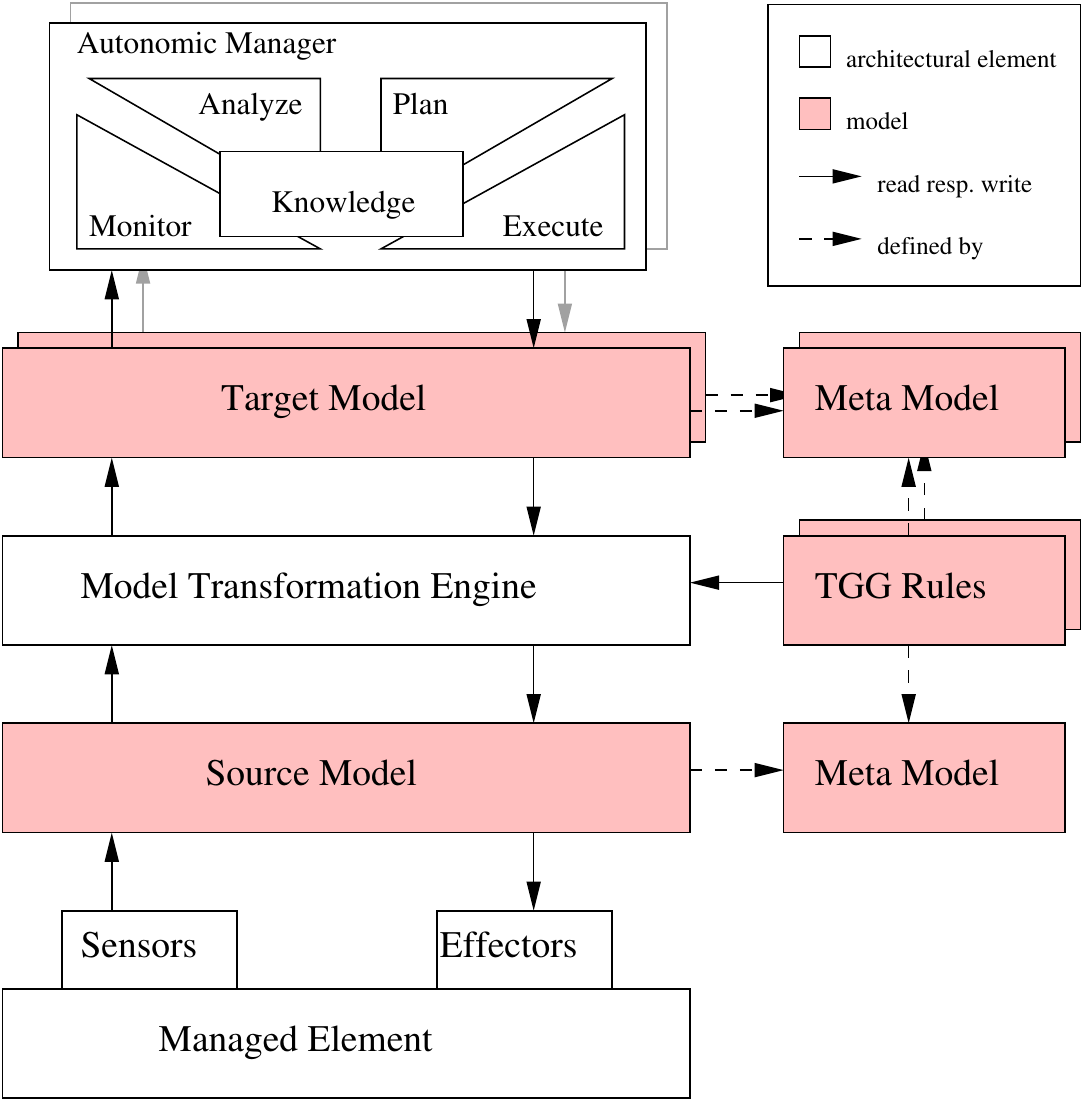}
        }
        \caption{Generic Architecture}
        \label{fig:gen-architecture}
        \vspace{-0.6cm}
\end{figure}

\noindent
Our approach combines \emph{Model-driven Engineering} (MDE) and the vision of autonomic computing. MDE techniques are employed to provide models describing different views on a managed software systems for different control loops being concerned with a certain self-management capability. 
The generic architecture of our approach is depicted in Figure~\ref{fig:gen-architecture}. It extends the control loop as proposed in \cite{McKinley+2004} by introducing models as the interface between \emph{Autonomic Managers} and the \emph{Managed Element}.
\emph{Sensors} and \emph{Effectors} provide a model-based view on a managed system in the form of a \emph{Source Model} that can be used for monitoring or adapting the system at run-time. A source model represents all capabilities of sensors and effectors. Consequently, a source model might be quite complex, which makes it laborious to use it as a basis for autonomic managers implementing the control loop activities monitoring, analysis, planning and execution. Thus, we propose using model transformation techniques to derive several \emph{Target Models} at run-time. Each target model raises the level of abstraction w.r.t.~the source model and it provides a specific view on the software system required for an autonomic manager focusing on a certain self-management capability. E.g., a manager being concerned with \emph{self-optimization} uses only those specific models that describe the resource utilization and performance attributes of a system, but does not have to consider views that are covered by other managers focusing on \emph{self-healing} or \emph{self-protection}.
Consequently, several autonomic managers work on possibly different target models as depicted at the top of Figure~\ref{fig:gen-architecture}.

The different target models are maintained by our generic \emph{Model Transformation Engine} being based on \emph{Triple Graph Grammars} (TGG)~\cite{GieseHildebrandt08,GW2008}. Source and target models are causally connected, i.e., changes in a source model are reflected in target models (monitoring) and vice versa (adaptation). This is possible due to the bidirectional transformation and synchronization capabilities of TGGs present in our engine, which work incrementally facilitating an efficient application at run-time. How two models have to be synchronized is specified declaritvely by \emph{TGG Rules} at the level of \emph{Meta Models} for the source and target models. Hence, the rules are independent of concrete models. Synchronization between source and target models can be triggered on demand enabling concurrent operations of managers and the decoupling of target models from a source model to ensure consistent analysis or to transfer several target model changes to the source model atomically. 

The implementation of our approach currently covers the monitoring capabilities. It is based on sensors the AC infrastructure \emph{mKernel}~\cite{bruhn2008Comprehensive} provides for monitoring systems being realized with \emph{Enterprise Java Beans 3.0} (EJB)~\cite{jsr220} technology. Based on the capabilities of these sensors, we developed a meta model for the EJB domain that defines the source model. For target models we developed three meta models and the required TGG rules covering the architecture, performance and failure states, which aim at autonomic managers being concerned with self-configuration, self-optimization, and self-healing, respectively. Our model transformation engine implementation and all meta models are based on the \emph{Eclipse Modeling Framework}. However, the whole infrastructure can run decoupled from the \emph{Eclipse} workbench.

\section{Conclusion \& Future Work}\noindent
We have presented an approach to support the architectural monitoring and adaptation by using meta models and model transformation techniques that operate efficiently and online and that can address different self-management capabilities.
As our current solution fully automates the monitoring of a system, we plan to cover the adaptation of architectures and to investigate the degree of automation for adaptations. 


\end{document}